\documentclass{iau} 
\usepackage{graphicx}

\title[22 Year Solar Magnetic Cycle and the Convection Zone Dynamics] 
{22 Year Solar Magnetic Cycle and its relation to Convection Zone Dynamics}

\author[Jain et al. ] 
{Kiran Jain$^1$, Sushanta Tripathy$^1$, Rudolf Komm$^1$, Frank Hill$^1$ and  Rosaria Simoniello$^2$}

\affiliation{$^1$National Solar Observatory, 3665 Discovery Drive, Boulder, CO 80303, USA\\ 
email: {\tt kjain@nso.edu}, {\tt stripathy@nso.edu},  {\tt rkomm@nso.edu},{\tt fhill@nso.edu} \\[\affilskip]
$^2$Geneva Observatory, University of Geneva, Geneva, Switzerland \\
email: {\tt rosaria.simoniello@unige.ch}}

\pubyear{2018}
\volume{340}  
\jname{Long-Term Datasets for the Understanding of Solar and Stellar Magnetic Cycle}
\editors{eds.}
\begin{document}

\maketitle

\begin{abstract}
Using continuous observations for 22 years from ground-based network GONG and 
space-borne instruments MDI onboard {\it SoHO} and  HMI onboard {\it SDO}, we
report both global and 
local properties of the convection zone and their variations with time.
\end{abstract}

\section{Introduction}
The dynamics of the convection zone plays a crucial role in understanding the activity cycles 
and to predict the strength of the next solar cycle. The solar dynamo, 
which  governs the solar activity, is believed to be seated in a thin layer 
called the {\it tachocline} located at the 
base of the convection zone. It can only be studied by using 
the techniques of helioseismology, where propagating acoustic waves are used to infer the 
properties of the region they travel through. The availability of continuous Dopplergrams at high-resolution 
and high-cadence allow us to map  this region over a 
complete Hale magnetic cycle. Here, we present some important results from our ongoing investigations.

\section{Data and Results}

The data used here are from a ground-based network, Global Oscillation
Network Group (GONG), and the space-borne instruments Michelson Doppler Imager (HMI) onboard {\it SoHO} and 
Helioseismic and Magnetic Imager (HMI) onboard {\it SDO}. 

\begin{figure}
 \hspace*{1.0 cm}
\begin{center}
{
 
   \includegraphics[width=4.5in]{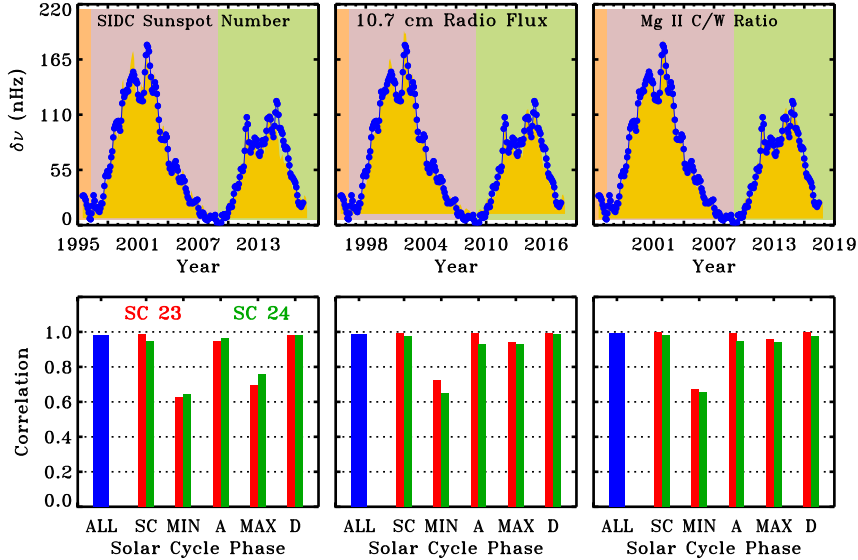} 
 }
  \caption{(Upper panels) Temporal variation of frequency shifts ($\delta\nu$)
  for common modes in all GONG data sets. Errors in calculated shifts are smaller than the size of 
  the symbols. Filled areas show the variation of solar activity indices.
  (Lower panels) The Pearson's linear correlation  for different phases: 
  minimum (MIN), ascending (A), maximum (MAX), and descending (D).}
   \label{fig1}
\end{center}
\end{figure}

The oscillation frequencies vary in phase with the solar activity cycle and display a strong correlation 
between them. However, un-interrupted observations for complete solar cycles suggest 
that the correlation changes  when the individual phases of the solar cycle 
are considered (Figure~\ref{fig1}), in particular it decreases significantly during the minimum phase (\cite[Jain et al. 2009]{Jainetal09}).
In addition to the variation on a solar-cycle time scale, the oscillation frequencies also have a
periodicity of approximately two years (\cite[Simoniello et al. 2013]{Simonielloetal13}). 
Further, similar to the magnetic activity on the solar surface, the oscillation frequencies 
also display the hemispheric asymmetry  (\cite[Tripathy et al. 2015]{Tripathyetal15})
as well as the latitudinal progression of the
solar cycle in the convection zone (\cite[Simoniello et al. 2016]{Simonielloetal16}).

The Dopplergrams are also used to map the flows in convection zone at different latitudes
as well as hemispheres. The zonal flow has bands of faster and slower rotations, the patterns commonly known
as the torsional oscillations (\cite[Komm et al. 2014]{Kommetal14}). The meridional flows are poleward with a peak amplitude
of about 16 to 20 m/s depending on the depth at about 40$^{\circ}$ latitude (\cite[Komm et al. 2015]{Kommetal15}).
These continuous observations further allow us to study the subsurface characteristics of active regions (ARs) (e.g., 
\cite[Jain et al. 2017]{Jainetal17}, and references therein). Using GONG Dopplergrams,
we have investigated the plasma flow beneath AR 10486 and AR 12192.
These were the biggest ARs of their respective solar cycles and produced several high M-
and X-class flares but with different CME productivity. We find that these ARs had unusually 
large horizontal flow amplitude with distinctly different directions. While meridional flow in AR 
12192 was poleward, it was equatorward in AR 10486 
(\cite[Jain et al. 2017]{Jainetal17}).  The flow patterns produced strong 
twists in horizontal velocity with depth in AR 10486 that persisted throughout the disk passage, 
as opposed to AR 12192, which produced a twist only after the eruption of the X3.1 flare that disappeared soon after.

In summary, the observations spanning two solar cycles have significantly advanced our understanding
of the different layers below the surface. However, there are still many areas where the
results are inconclusive due to large discrepancies between different methods and also between
data from different instruments. Hence, long-term observations for many more cycles are 
essential to fully understand the dynamics of the convection zone.

Data were acquired by GONG instruments operated by NISP/NSO/AURA/NSF. {\it SoHO} is a mission of international cooperation
 between ESA and NASA. {\it SDO} data courtesy of {\it SDO} (NASA) and the HMI and AIA consortium.


\begin{thebibliography}{}


\bibitem[Jain et al. 2009]{Jainetal09}
{Jain, K., Tripathy, S.C., \& Hill, F.},
2009, \textit{ApJ}, 695, 1567

\bibitem[Jain et al. 2017]{Jainetal17}
{Jain, K., Tripathy, S.C., \& Hill, F.},
2017, \textit{ApJ}, 849, 94

\bibitem[Komm et al. 2015]{Kommetal15}
{Komm, R., Gonz{\'a}lez Hern{\'a}ndez, I., Howe, R., \& Hill, F.},
2015, \textit{Sol. Phys.}, 290, 3137

\bibitem[Komm et al. 2014]{Kommetal14}
{Komm, R., Howe, R., Gonz{\'a}lez Hern{\'a}ndez, I., \& Hill, F.},
2014, \textit{Sol. Phys.}, 289, 3435

\bibitem[Simoniello et al. 2013]{Simonielloetal13}
{Simoniello, R., Jain, K., Tripathy, S.C. et al.},
2013, \textit{ApJ}, 765, 100

\bibitem[Simoniello et al. 2016]{Simonielloetal16}
{Simoniello, R., Tripathy, S.C., Jain, K., \& Hill, F.},
2016, \textit{ApJ}, 828, 41

\bibitem[Tripathy et al. 2015]{Tripathyetal15}
{Tripathy, S.C., Jain, K., \& Hill, F.},
2017, \textit{ApJ}, 812, 20

\end{thebibliography}
\end{document}